\documentclass[reprint,nofootinbib,notitlepage,onecolumn]{revtex4-2}

\usepackage{amsfonts}
\usepackage{graphicx}
\usepackage{bm}
\usepackage{amssymb,amsmath}
\usepackage{mathtools}
\usepackage[utf8]{inputenc}

\begin{document}

\title{The massive side of the electromagnetic waves}
\author{Rafael Ferraro}
\email{ferraro@iafe.uba.ar}
\thanks{member of Carrera del Investigador Cient\'{\i}fico (CONICET,
Argentina)} \affiliation{Instituto de Astronom\'{\i}a y F\'{\i}sica del
Espacio (IAFE, CONICET-UBA), Buenos Aires, Argentina \\ Departamento
de F\'{\i}sica, Facultad de Ciencias Exactas y Naturales, Universidad de
Buenos Aires, Argentina}

\begin{abstract}
Stationary electromagnetic waves display aspects that are shared
with massive particles, since the energy and momentum contained in a
volume of sides equal to the wavelengths form a non-zero
energy-momentum invariant. The parallel can be extended to the
notion of weight when this concept makes sense, that is, in the
Newtonian chart of a weak inertial-gravitational field.
\end{abstract}

\maketitle

\section{Introduction}
Particles and fields exhibit their energy-momenta through different
geometric objects. Particle energy-momentum is represented by a
vector $p^{\mu}$ composed of energy, $p^0=c^{-1}E$, and momentum
$p^i$ ($i=1,2,3$); the momentum is the energy flux since
$p^i=c^{-2}E\, dx^i/dt$. Instead in continuous systems, such as
fluids and fields, the energy and momentum densities together with
their respective fluxes, are captured by the energy-momentum tensor
$T^{\mu\nu}$: $T^{00}$ is the energy density, $c^{-1} T^{i0}$ is the
momentum density, and $cT^{0i}$ is the energy per unit of traversed
area and per unit of time. Besides the energy-momentum tensor
contains the stresses $T^{ij}$ between adjacent elements of the
continuous system (momentum fluxes).

The state of rest of a continuous system is defined by the condition
$T^{0i}=0=T^{i0}$. Therefore, if a boost $z^{\prime }=\gamma
(z+\beta x^{0})$, $x^{0^{\prime }}=\gamma (x^{0}+\beta z)$ is
performed starting from the frame where the fluid is at rest, then
the transformed energy and momentum densities will become
\begin{equation}
T^{0^{\prime }0^{\prime }}=\frac{\partial x^{0\prime }}{\partial x^{\mu }}%
\frac{\partial x^{0\prime }}{\partial x^{\nu }}~T^{\mu \nu }=\gamma
^{2}~(T^{00}+\beta ^{2}~T^{zz})~,  \label{energy}
\end{equation}%
\begin{equation}
T^{z^{\prime }0^{\prime }}=\frac{\partial z^{\prime }}{\partial x^{\mu }}%
\frac{\partial x^{0\prime }}{\partial x^{\nu }}~T^{\mu \nu }=\gamma
^{2}~\beta ~(T^{00}+T^{zz})~,  \label{momentum}
\end{equation}%
\begin{equation}
T^{x^{\prime }0^{\prime }}=\frac{\partial x^{\prime }}{\partial x^{\nu }}%
\frac{\partial x^{0\prime }}{\partial x^{\mu }}~T^{\mu \nu }=\gamma
\beta
~T^{xz}~,~~\ \ \ \ \ ~~~T^{y^{\prime }0^{\prime }}=\frac{\partial y^{\prime }%
}{\partial x^{\nu }}\frac{\partial x^{0\prime }}{\partial x^{\mu
}}~T^{\mu \nu }=\gamma \beta ~T^{yz}~.  \label{flux}
\end{equation}%
As can be seen, the stresses $T^{ij}$ enter the transformations
producing momentum not only in the direction of the boost, but in
the other directions too. This behavior is proper of a tensor, and
is quite different from that of a particle; the transformation of
the energy-momentum of a particle starting from its proper frame is
instead
\begin{equation}
p^{0^{\prime }}=\gamma ~p^{0},~~\ \ \ \ ~\ ~\ \ ~\ p^{z^{\prime
}}=\gamma \beta ~p^{0},~~\ \ \ \ ~\ ~\ \ ~\ p^{x^{\prime }}=0~,~~\ \
\ \ ~\ ~\ \ ~\ p^{y^{\prime }}=0~.\label{pp}
\end{equation}%
If the shear stresses $T^{xz}$, $T^{yz}$ are zero, then the sole
difference between the continuous system and particles will be the
contribution of the pressure $T^{zz}$ in Eqs.~(\ref{energy}) and
(\ref{momentum}).\footnote{In addition, the $\gamma$ factor appears
squared in the transformation for continuous systems. This is
consistent with the character of density of the components of
$T^{\mu \nu }$, since the volume is contracted in the boost
direction.} If, for instance, the system is a non-relativistic
perfect fluid the pressure $T^{zz}$ can be neglected if compared
with the \textit{rest energy} density $T^{00}$. Because of this
reason, a non-relativistic perfect fluid behaves like an aggregate
of particles in interaction. On the contrary, if the fluid were a
photon gas, then $T^{zz}$ could not be ignored; since the trace of
the electromagnetic energy-momentum tensor is identically zero, its
diagonal components, such as $T^{00}$ and $T^{zz}$, are comparable.
In particular, for a photon gas in Minkowski spacetime it must be
$T^{xx}=T^{yy}=T^{zz}=T^{00}/3$, as dictated by the isotropy (so,
the state equation of the gas, which relates the rest or proper
pressure $p$ to the rest energy density $\rho$, is $p=\rho /3$).
Therefore Eq.~(\ref{momentum}) becomes $T^{z^{\prime }0^{\prime
}}=(4/3)~\gamma ^{2}~\beta ~T^{00}$. The factor $4/3$ could be
unexpected in some way, but it is not exclusive of the photon gas.
The factor $4/3$ also appears in the (isotropic) case of the
pointlike charge, once the energy and momentum densities are
integrated in the space (see Ref.~\onlinecite{Feynman}, \S 28.2).

\bigskip

In sum, the continuous systems that display a particle-like
dynamical behavior are those that lack contributions to the stress
tensor along the direction of the motion in the proper frame where
the system is at rest; according to Eqs.~(\ref{energy}-\ref{flux})
it must be $T^{iz}=0$. In the following sections we will show that
there are electromagnetic configurations that display a
particle-like behavior, not only in the fulfillment of
transformations of the type (\ref{pp}) but also in the relationship
between rest energy and weight. This behavior will emerge when the
respective densities are integrated in a box characterized by the
wavelengths and the period.

\section{Electromagnetic field in a waveguide}\label{waveguide}
In Minkowski spacetime let us consider the electromagnetic potential
$A_{\mu }=A~(0,0,0,1)$ where
\begin{equation}
A=\frac{E_{o}}{\omega }~\cos \omega t ~\sin k x~\sin K y~,~\
~~~~~~\omega =c\sqrt{k^{2}+K^{2}}~.\label{potential}
\end{equation}
This potential is a trivial solution to Maxwell equations. Its field
$F_{\mu \nu }=\partial _{\mu }A_{\nu }-\partial _{\nu }A_{\mu }$
satisfies appropriate boundary conditions for a waveguide having a
rectangular section in the $x-y$ plane with $0<x<m\pi /k$ and
$0<y<n\pi /K$ ($m,n\in N$). The solution corresponds to a transverse
magnetic mode (TM) since the magnetic field is perpendicular to the
$z-$axis \cite{Jackson}.

The electromagnetic energy-momentum tensor $ T^{\mu \nu }$ for the
metric signature $(+---)$ is
\begin{equation}
T^{\mu \nu }=\frac{1}{\mu _{o}}~\left( F_{~\lambda }^{\mu
}F^{\lambda \nu }- \frac{1}{4}~g^{\mu \nu }~F^{\rho \lambda
}F_{\lambda \rho }\right) ~.
\end{equation}
No energy propagates along the waveguide because $T^{0z}=0=T^{z0}$.
The electromagnetic waves bounce between the walls of the waveguide
traveling in the $x-y$ plane, so giving rise to stationary waves.
This means that $T^{x0}$, $T^{y0}$ are not zero; however, when
averaged on a period $\tau =2\pi \omega ^{-1}$, or when integrated
on the waveguide section, they are zero. So it can be stated that
the configuration (\ref{potential}) is ``at rest''. The \textit{rest
energy} $\mathcal{E}_{o}$ is obtained by integrating the density
$T^{00}$ in the space; in particular, in the cell whose sides
$\Delta x$, $\Delta y$ are equal to the respective wavelengths, and
$\Delta z$ is arbitrary, the energy is
\begin{equation}
\mathcal{E}_{o}=\int_{0}^{2\pi /k}\!\int_{0}^{2\pi
/K}\!\int_{z}^{z+\Delta z}dx\,dy\,dz~T^{00}~=~\frac{E_{o}^{2}~\pi
^{2}}{2~\mu _{o}~c^{2}kK}~\Delta z~=~\frac{E_{o}^{2}}{8~\mu
_{o}~c^{2}}\times\textrm{volume}~.  \label{Erest}
\end{equation}
$\mathcal{E}_{o}$ does not depend on time $t$, as should be expected
since $\partial /\partial t$ is a Killing vector.

In a frame uniformly moving along the waveguide, the energy
$\mathcal{E}^{\prime }$ is obtained by integrating $T^{0^{\prime
}0^{\prime }}$ at a constant $t^{\prime }$. Since the phase is
Lorentz-invariant, it is
\begin{equation}
\omega t=\omega ~\gamma (t^{\prime }-c^{-1}\beta ~z^{\prime })
\end{equation}
Then it results
\begin{equation}
\omega ^{\prime }=\omega ~\gamma ~,~~\ \ \ \ ~k_{z^{\prime
}}=c^{-1}\omega ~\gamma \beta ~\ ~~~\Longrightarrow ~~\ ~~~~\omega
^{\prime ~2}-k_{z^{\prime }}^{2}~c^{2}=\omega ^{2}~.
\label{dispersion}
\end{equation}%
According to Eq.~(\ref{energy}), the energy $\mathcal{E}^{\prime }$
in a cell is
\begin{equation}
\mathcal{E}^{\prime }=\int_{0}^{2\pi /k}\!\int_{0}^{2\pi
/K}\!\int_{0}^{2\pi /k_{z^{\prime }}}dx\, dy\, dz^{\prime }~\gamma
^{2}~(T^{00}+\beta ^{2}~T^{zz})|_{t=\gamma (t^{\prime }-c^{-1}\beta
z^{\prime })}~, \label{intEprime}
\end{equation}
($x$, $y$ are invariant under a boost in the $z$ direction), which
turns out to be
\begin{equation}
\mathcal{E}^{\prime }=\gamma ~\frac{E_{o}^{2}~\pi ^{2}}{2~\mu _{o}~c^{2}kK}~%
\frac{2\pi \gamma }{k_{z^{\prime }}}~.  \label{Eprime}
\end{equation}
By comparing this result with Eq.~(\ref{Erest}) we realize that the
particle-like relationship between the rest energy $\mathcal{E}_{o}$
and $\mathcal{E}^{\prime }$ is got by understanding $\Delta z$ in
(\ref{Erest}) as the proper length $\gamma \lambda _{z^{\prime }}$
associated with $\lambda _{z^{\prime }}=2\pi /k_{z^{\prime
}}$;\footnote{If $\beta \rightarrow 0$, then $k_{z^{\prime }}$ goes
to zero in Eq.~(\ref{dispersion}); so $\Delta z$ diverges. This just
says that the total energy in the (infinite) waveguide is infinite.}
\begin{equation}
\mathcal{E}^{\prime }=\gamma ~\mathcal{E}_{o}~.\label{transfE}
\end{equation}
On the other hand, the momentum of the cell is directed along the
waveguide.\footnote{The momentum components $x$, $y$ results from
integrating the Eq.~(\ref{flux}). They are zero, as the symmetry
anticipates.} According to Eq.~(\ref{momentum}) the momentum of the
cell is
\begin{eqnarray}
\mathcal{P}^{\prime }&=&\int_{0}^{2\pi /k}\!\int_{0}^{2\pi
/K}\!\int_{0}^{2\pi /k_{z^{\prime }}}dx\, dy\, dz^{\prime
}~c^{-1}~T^{z^{\prime }0^{\prime }}\label{intPprime} \\
&=&\int_{0}^{2\pi /k}\!\int_{0}^{2\pi /K}\!\int_{0}^{2\pi
/k_{z^{\prime }}}dx\, dy\, dz^{\prime }~c^{-1}\gamma ^{2}\beta
~(T^{00}+T^{zz})|_{t=\gamma (t^{\prime }-c^{-1}\beta z^{\prime })}~,
\notag
\end{eqnarray}
which turns out to be
\begin{equation}
\mathcal{P}^{\prime }=c^{-1}\gamma \beta ~\frac{E_{o}^{2}~\pi
^{2}}{2~\mu
_{o}~c^{2}kK}~\frac{2\pi \gamma }{k_{z^{\prime }}}=c^{-1}\gamma \beta ~%
\mathcal{E}_{o}  \label{Pprime}
\end{equation}
(it does not depend on $t$ because $\partial /\partial z$ is a
Killing vector). So $\mathcal{P}^{\prime }$ has the form of a
particle momentum: it satisfies
\begin{equation}
\mathcal{E}^{\prime }{}^{2}-\mathcal{P}^{\prime
}{}^{2}c^{2}=\mathcal{E}_{o}^{2}\ ,  \label{invariant}
\end{equation}
which reproduces the dispersion relation (\ref{dispersion}).
Noticeably, the presence of $T^{zz}$ in Eqs.~(\ref{intEprime}) and
(\ref{intPprime}) does not interfere with the particle-like behavior
because the integral of $T^{zz}$ in Eq.~(\ref{intPprime}) is zero.
The result shows that the electromagnetic configuration
(\ref{potential}) can be regarded as an aggregate of cells
displaying a particle-like behavior. In fact, the energy-momentum of
each cell is contained in the four-vector
$\mathcal{P}^\mu=(c^{-1}\mathcal{E},0,0,\mathcal{P})$, whose
(invariant) squared norm $c^{-2}\mathcal{E}_o^2$ represents the
squared mass (times $c^2$) of the cell. The dimensions of the cells
are given by the wavelengths. Neighboring cells interact through
pressures and tensions; no shear stresses appear (they are zero at
the nodes of the stationary wave; besides, the integrals of $T^{xz}$
and $T^{yz}$ in the cell section are zero). In the next sections we
will show that this particle-like behavior of cells extends also to
the relation between rest energy and weight, whenever the notion of
weight can be properly introduced.

\section{The weight of the energy}
The equality between gravitational and inertial masses is one of the
cornerstones of general relativity. It implies that inertial and
gravitational forces make up a single inertial-gravitational field
affecting all particles and becoming locally zero in a suitable
chart; this is the content of the weak equivalence principle. The
inertial-gravitational field defines the spacetime geometry and
characterizes the geodesic curves; in particular, the timelike
geodesics are the worldlines of the universal free fall particle
motion. The implementation of Einstein's equivalence principle is
made through the minimal coupling prescription, which states that
the laws we use in special relativity are translated into general
relativity by replacing partial derivatives $\partial_\nu$ with the
covariant derivatives $\nabla_\nu$ of the Levi-Civita affine
connection. The \textit{locally inertial frames}, where the laws
recover the form they have in special relativity (i.e., the
inertial-gravitational field becomes locally zero), are those where
the metric tensor $g_{\mu\nu}$ locally has the form $\eta_{\mu \nu
}=\mathrm{diag}\{1,-1,-1,-1\}$ and its first partial derivatives
become zero (i.e., the Levi-Civita connection is locally zero). The
charts fulfilling these conditions are related to each other through
Lorentz transformations.

Although the inertial-gravitational field is a single field, under
the \textit{weak field approximation} it is possible to choose
\textit{Newtonian charts} where the dynamical equations of free fall
particles read, like in Newton's theory, as a balance between an
inertial term and a gravitational term associated with the particle
weight $mg$. In fact the free fall equations, as dictated by the
minimal coupling prescription, state that the covariant derivative
of the energy-momentum is zero,
\begin{equation}
\frac{Dp^{\mu }}{D\tau }=u^{\nu }\nabla _{\nu }p^{\mu }=0~,
\end{equation}%
where $p^{\mu }=mu^{\mu }$ is the particle energy-momentum, $u^{\mu
}=dx^{\mu }/d\tau $ is its four-velocity and $\tau $ is its proper
time. The Levi-Civita connection matches this equation with the
geodesic equation for the metric $g_{\mu\nu}$.

Let us consider a particle instantaneously at rest. This can be
achieved by a suitable choice of the chart, and allows to avoid
contributions of the particle motion to the connection terms. Then
it is $u^{\mu} = (c~g_{00}^{-1/2},0,0,0)$, so the free fall
equations turn out to be
\begin{equation}
\frac{du^{\mu }}{d\tau }+\Gamma _{00}^{\mu }~g_{00}^{-1}~c^{2}=0~.
\end{equation}
If the geometry is static we can choose a static chart where the
metric components $g_{\mu \nu }$ do not depend on $x^{0}$, and the
components $g_{0i}$ are zero ($i=1,2,3$). Then the Levi-Civita
connection turns out to be $\Gamma _{00}^{0}=0$ and $\Gamma
_{00}^{i}=-1/2~g^{ij}g_{00,j}$; thus the previous equations become
\begin{equation}
\frac{du^{0}}{d\tau }=0~,~~~~\ ~\frac{du^{i}}{d\tau }=\frac{1}{2}%
~g^{ij}~g_{00,j}~g_{00}^{-1}~c^{2}~.~
\end{equation}%
These equations are valid in any static chart. In a locally inertial chart $%
g_{00,j}$ locally vanishes, so the motion is instantaneously
uniform. Instead in a Newtonian chart $g_{00}$ relates to the
Newtonian gravitational potential $\Phi $ (potential energy per unit
of mass) according to $g_{00}\simeq 1+2c^{-2}\Phi$
\cite{Schutz,Carroll,Ferraro}; the weak field approximation implies
that $c^{-2}|\Phi |<<1$. In this chart we recover the Newtonian
dynamics
\begin{equation}
\frac{du^{0}}{d\tau }=0~,~~~~\ ~\frac{du^{i}}{d\tau }\simeq -\delta
^{ij}~\Phi _{,j}~.
\end{equation}
For a uniform gravitational field it is $\Phi =gz$, then at the
lowest order in $c^{-2}gz$ one obtains
\begin{equation}
\frac{d^{2}x^{0}}{d\tau ^{2}}=0~,~~~~\ ~\frac{d^{2}x}{d\tau
^{2}}\simeq
0~,~~~~\ ~\frac{d^{2}y}{d\tau ^{2}}\simeq 0~,~~~~\ ~\frac{d^{2}z}{d\tau ^{2}}%
\simeq -g~.  \label{gparticle}
\end{equation}
The first equation says that the relation between $x^{0}$ and $\tau
$ is linear at the considered instant. Therefore, the last equation
means that the acceleration measured in terms of $z$ and $x^{0}$ is
equal to $-g$ like in classical physics. Thus $(0,0,0,-mg)$ plays,
in the Newtonian chart, the role of the weight of the particle.

Since $m$ is the rest energy times $c^{-2}$, we are aimed to find
out whether in continuous systems the weight is also expressed as
the rest energy times $c^{-2}g$. To do this, we will repeat the
previous strategy which consists in writing the dynamical equations
in conditions where the momentum is null (system at rest), and then
using the Newtonian chart to analyze the way the inertial terms
separate from the gravitational terms coming from the connection.

\subsection{The perfect fluid}
In continuous systems, the energy-momentum tensor $T^{\mu \nu }$
takes the role that $p^{\mu }$ plays for a particle. Then, the free
fall equations $u^{\nu }\nabla _{\nu }p^{\mu }=0$ are replaced with
$\nabla _{\nu }T^{\mu \nu }=0$. Taking into account that $\Gamma
_{\nu \lambda }^{\lambda }=1/2~g^{\lambda \rho }~\partial _{\nu
}g_{\lambda \rho }=\partial _{\nu }\ln \sqrt{|\det \mathbf{g}|}$,
then the equation~$\nabla _{\nu }T^{\mu \nu }=0$ turns out to be
\begin{equation}
0=\nabla _{\nu }T^{\mu \nu }=|\det \mathbf{g}|^{-1/2~}\partial _{\nu
}\left( |\det \mathbf{g}|^{1/2~}T^{\mu \nu }\right) +\Gamma
_{\lambda \rho }^{\mu }~T^{\lambda \rho }~.  \label{fluid}
\end{equation}

To exemplify let us consider a perfect fluid, whose energy-momentum
tensor is
\begin{equation}
T^{\mu \nu }=(\rho +p)~c^{-2}~u^{\mu }u^{\nu }-p~g^{\mu \nu }~.
\end{equation}
For a static chart and a fluid locally at rest (i.e., $u^{\nu
}=c~(g_{00}^{-1/2},0,0,0)$) it results%
\begin{equation}
T^{00}=g_{00}^{-1}~\rho
~,~~~~~~~~T^{0i}=0=T^{i0}~,~~~~~~~~T^{ij}=-p~g^{ij}~, \label{diag}
\end{equation}
which shows that the energy flux and the momentum density are zero.
If in addition the chart is locally inertial, then it is $g_{00}=1$,
$g^{ij}=-\delta ^{ij}$; thus we see that $\rho $ is the rest energy
density and $p$ is the pressure as defined in special relativity.
Besides, the perfect fluid does not present shear stresses; it is
isotropic.

On the other hand, to study the fluid in the Newtonian chart where
$g_{00}\simeq 1+2c^{-2}\Phi $, we have to make the rest of the
components $g_{\mu \nu }$\ explicit. For the potential $\Phi =gz$,
we will use
\begin{equation}
g_{\mu \nu }=\mathrm{diag}\{1+2c^{-2}gz,-1,-1,-(1+2c^{-2}gz)^{-1}
\}~,~~~~\ \ ~~~~|\det \mathbf{g}|=1~,  \label{weak}
\end{equation}
which is a vacuum solution to Einstein equations. As can be seen,
the coordinates $x$, $y$ keep the meaning of physical distances
because no gravitational effects exist in these directions.

If Eqs.~(\ref{diag}) are satisfied in the Newtonian chart (i.e., if
the fluid is at rest in the uniform inertial-gravitational field),
then Eq.~(\ref{fluid}) for $\mu =z$ is
\begin{equation}
0=\partial _{z~}T^{zz}+\Gamma _{00}^{z}~g_{00}^{-1}~\rho
-\sum_{j=1}^{3}~\Gamma _{jj}^{z}~g^{jj}~p~,  \label{Tcov}
\end{equation}
since the diagonality of $T^{\mu \nu }$ in Eq.~(\ref{diag}) implies
that $\partial_{z} $ is the only derivative that appears. The
non-zero Christoffel symbols for the metric (\ref{weak}) are
\begin{equation}
\Gamma _{0z}^{0}=\Gamma _{z0}^{0}=-\Gamma _{zz}^{z}=\frac{c^{-2}~g}{%
1+2c^{-2}gz}~,~~\ ~~\Gamma _{00}^{z}=c^{-2}~g~(1+2c^{-2}gz)~.
\label{Christoffel}
\end{equation}
Thus Eq.~(\ref{Tcov}) results in
\begin{equation}
0=(1+2c^{-2}gz)~\partial_{z}p+c^{-2}~g~(\rho +p)~,
\end{equation}
which at the lowest order is
\begin{equation}
0\simeq \partial _{z~}p+c^{-2}g~(\rho + p)~.  \label{hydrostatics}
\end{equation}
This hydrostatic equation shows how the pressure must vary with $z$
to support the weight of the fluid. As we can see, the rest energy
density in the second term is accompanied by the pressure, which is
a typical relativistic result. In most cases, the pressure is
negligible compared to the rest energy density; so we obtain that
the pressure gradient balances the weight per unit of volume
$c^{-2}\rho ~g=\rho _{m}~g$ ($\rho _{m}$ is the mass density), and
the equation is trivially integrated for a non-compressible fluid to
give the classical result
\begin{equation}
p+c^{-2}\rho ~g~z\simeq constant~.
\end{equation}
In sum, the weight of a perfect fluid per unit of volume has the
form $c^{-2}\rho g z$ whenever it is possible to ignore the pressure
with respect to the rest energy density. In the following section we
will study the case of the electromagnetic field in a waveguide,
whose $T^{\mu \nu }$ is not diagonal, and whose pressures and shear
stresses cannot be neglected.

\section{Stationary configuration in the presence of the uniform
inertial-gravitational field} Let us now come back to the
electromagnetic configuration of Section \ref{waveguide}. To
complete the analogy between particles and cells in the waveguide,
we will pass to consider the weight of the cells. The literature on
waveguides in gravitational fields is rather sparse; let us mention
here Ref.~\cite{Beig}.\footnote{Waveguide configurations can also be
considered for gravitational waves. For instance, the propagation of
a gravitational wave in the (external) field of a filament of
galaxies has been studied in Ref.~\cite{Bimonte}.} Our aim is to
reconsider the stationary electromagnetic configuration of Section
\ref{waveguide} now in the presence of a weak uniform gravitational
field directed along the waveguide. For this purpose we will solve
Maxwell equations $\partial _{\mu }(|\det \mathbf{g}|^{1/2~}F^{\mu
\nu })=0$ in the metric (\ref{weak}). It is not difficult to verify
that the corresponding stationary solution at the first order in
$gc^{-2}z$ can be written as \footnote{For terrestrial gravity
$c^{2}/g\sim 10^{16}$m. So the main restriction is not
$|z|<<c^{2}/g$ but the approximation of uniformity.}
\begin{equation}
A_{\mu }\simeq \mathcal{A}~(~0,~gc^{-2} x,~gc^{-2}
y,~1)\label{gpotential1}
\end{equation}
where
\begin{equation}\label{gpotential2}
\mathcal{A}\simeq \frac{E_{o}~}{\omega }~(1-2~gc^{-2}z)~\cos \omega
t ~\sin [(1-gc^{-2}z) k x]~\sin [(1-gc^{-2}z) K y]~,
\end{equation}
and $\omega =c \sqrt{k^{2}+K^{2}}$.

So, the inertial-gravitational field affects the stationary
configuration in different ways; the amplitude and the wave numbers
become decreasing with $z$, and the potential acquires components in
$x$ and $y$. The wavelengths, that determine the sides of the cells,
are now
\begin{eqnarray}
\lambda _{x}(z)&=&\frac{2\pi }{(1-gc^{-2}z)~k}\simeq \frac{2\pi
}{k}\, (1+gc^{-2}z)~,\nonumber\\ \lambda _{y}(z)&=&\frac{2\pi}
{(1-gc^{-2}z)~K}\simeq \frac{2\pi }{K}\,(1+gc^{-2}z)~
\end{eqnarray}%
The configuration (\ref{gpotential1})-(\ref{gpotential2}) is ``at
rest'' in the sense that the time averages of the momentum densities
$T^{i0}$ are zero.

To compute the weight of a cell, let us recall that the dynamics of
a continuous system in special relativity is governed by the
equation \cite{Schutz,Ferraro}
\begin{equation}
\partial _{\nu }T^{\mu \nu }=f_{ext}^{\mu }~,  \label{SR}
\end{equation}
where $f_{ext}^{\mu }$ is the external four-force per unit of
volume. In general relativity a freely-falling system satisfies
Eq.~(\ref{fluid}); in the Newtonian chart, where the metric and the
connection are the ones of Eqs.~(\ref{weak}) and (\ref{Christoffel})
respectively, the four componentes of Eq.~(\ref{fluid}) are
\begin{equation}
\partial _{\nu }T^{0\nu }=-2~\Gamma _{0z}^{0}~T^{0z}~,
\end{equation}
\begin{equation}
\partial _{\nu }T^{x\nu }=0=\partial _{\nu }T^{y\nu }
\end{equation}
\begin{equation}
\partial _{\nu }T^{z\nu }=-\Gamma _{00}^{z}T^{00}-\Gamma _{zz}^{z}T^{zz}
\end{equation}
The connection terms contain the information about the gravitational
``force''. They look like the external four-force per unit of volume
in Eq.~(\ref{SR}). Although by themselves they do not form a
four-vector (they are just a part of the four-vector $\nabla _{\nu
}T^{\mu \nu }$), the isolation of those terms containing $g$ in the
Newtonian chart allows shaping the idea of weight, just as it was
done with the particle in Eq.~(\ref{gparticle}). In this limited
sense, the four components of the gravitational force per unit of
volume (at the lowest order in $g$) are
\footnote{$f_{grav}^{0}\simeq 0$ because $T^{0z}$ is already of
order $g$; remember that $T^{0z}=0$ if $g=0$.}
\begin{equation}
\text{``}f_{grav}^{\mu }\text{''}\simeq
-c^{-2}g~(0,~0,~0,~T^{00}-T^{zz})
\end{equation}%
By integrating in the volume of a cell we obtain the ``four-force''
\begin{equation}
\text{``}K_{grav}^{\mu }\text{''}\simeq -c^{-2}g~\int_{0}^{\lambda
_{x}}\int_{0}^{\lambda _{y}}\int_{z}^{z+\Delta
z}dx\,dy\,dz~(0,~0,~0,~T^{00}-T^{zz})~. \label{dK}
\end{equation}%
At the considered order, the integrand can be computed for $g=0$.
According to Eq.~(\ref{Erest}), the first term is $-c^{-2}
g~\mathcal{E}_{o}$; on the other hand, the integral of $T^{zz}$ is
proportional to $\cos (2\omega t)$. Therefore, by time averaging in
a period one gets
\begin{equation}
<\text{``}K_{grav}^{\mu }\text{''}>\; \simeq\; (0,~0,~0,-c^{-2}
g~\mathcal{E}_{o})~, \label{K}
\end{equation}
which means that the (averaged) weight of the cell is the rest
energy $\mathcal{E}_{o}$ times $c^{-2}g$, as it happens with
particles.

\section{Conclusions}
The propagation of electromagnetic waves in waveguides of simply
connected sections \cite{Ferraro2,Collin}, like the rectangular
section we have considered, possesses features which are proper of
massive particles. This behavior is not expected a priori for a
system whose pressures are comparable to the energy density; it
emerges when the energy and momentum densities are integrated in
cells whose sides are determined by the wavelengths, because the
contributions from the stresses cancel out (see
Eq.~(\ref{invariant})). While the energy-momentum distribution of
the field is \textit{locally} described by the tensor $T^{\mu\nu}$,
the energy-momentum of a cell is described by a \textit{time-like}
four-vector $\mathcal{P}^\mu$, as for a massive particle. The
time-like character of $\mathcal{P}^\mu$ comes from the periodic
boundary conditions the waveguide imposes to the field, and the
concomitant dispersion relation (see Eqs.~(\ref{potential}) and
(\ref{dispersion})).\footnote{$\mathcal{P}^\mu\mathcal{P}_\mu$ is
zero for free electromagnetic plane waves; see Ref.
\cite{Sturrock}.} Thus, a similar result could be expected if there
were compact extra-dimensions instead of boundaries.

The parallel between cells and massive particles is reinforced when
it is noted that a stationary electromagnetic wave in a
gravitational background also exhibits a particle-like behavior with
respect to its weight. In fact, by solving Maxwell equations in a
uniform inertial-gravitational field at the lowest order in
$c^{-2}gz$, we have succeeded in showing that also the
(time-averaged) weight of the cell is entirely analogous to the one
of a particle; i.e., it is equal to the rest energy times $c^{-2}g$
(see Eq.~(\ref{K})). It must be emphasized that the idea of
\textit{weight} is not covariant (the ``weight'' is not a
four-force); it is a notion that just emerges when the free fall
equations, as written in a Newtonian chart, are separated into
inertial and gravitational terms, as we do for particles.

Despite appearances, Minkowski flat geometry was not abandoned when
we introduced the chart (\ref{weak}) to describe a uniform
inertial-gravitational field. This could have been expected because
the uniform gravitational field generates force (connection) but
does not generate tidal forces (curvature). Actually the Newtonian
chart of Eq.~(\ref{weak}) can be obtained from an inertial chart
$\{X^{\mu }\}$, where the metric is $\eta _{\mu \nu }$, through the
transformation~\footnote{Use $g_{\mu \nu }=\frac{\partial X^{\lambda
}}{\partial x^{\mu }}\frac{\partial X^{\rho }}{\partial x^{\nu
}}~\eta _{\lambda \rho }~$.}
\begin{eqnarray}
c T\ \equiv\ X^{0}\!\!  &=&\!\!
c^{2}g^{-1}\sqrt{1+2~c^{-2}g~z}~\sinh
(c^{-2}g~x^{0})~,\nonumber\\
Z\!\!  &=&\!\!  c^{2}g^{-1}~\sqrt{1+2~c^{-2}g~z}~\cosh
(c^{-2}g~x^{0})-c^{2}g^{-1}
\end{eqnarray}%
which is a variant of Rindler coordinate transformation
\cite{Carroll} (notice that the curves $z=constant$ are the
hyperbolas $Z^{2}-c^2 T^2=constant$). At the lowest order these
transformations are
\begin{equation}
X^{0}\simeq x^{0}~(1+c^{-2}g~z)~,~\ \ \ \ ~\ ~Z\simeq z+\frac{1}{2}%
~c^{-2} g~({x^0}\;^2-z^{2})~.
\end{equation}
Therefore, the coordinate lines $z=constant$ of the Newtonian chart
are uniformly accelerated worldlines $Z\simeq constant+g\; T^2/2$.
The acceleration is equal to $g$, in agreement with the equivalence
principle that states the indistinguishability between the
acceleration of the frame and the existence of a static and uniform
gravitational field. The configurations we characterized as ``at
rest'' in the uniform gravitational field were actually at rest in a
uniformly accelerated frame of Minkowski spacetime, in a full
realization of the equivalence principle. Certainly, for non-uniform
inertial-gravitational fields the equivalence is only local.

\vskip1cm

\begin{acknowledgments}
This work was supported by Consejo Nacional de Investigaciones
Cient\'{\i}ficas y T\'{e}cnicas (CONICET) and Universidad de Buenos Aires.
\end{acknowledgments}

\end{document}